# Terahertz generation by optical rectification in uniaxial birefringent crystals

**J. D. Rowley,[1] J. K. Wahlstrand,[2] K. T. Zawilski,[3] P. G. Schunemann,[3] N. C. Giles,[4] and A. D. Bristow[1,*]**

[1] *Department of Physics, West Virginia University, Morgantown, West Virginia, 26506-6315,USA*
[2] *Institute for Research in Electronics and Applied Physics, University of Maryland, College Park, Maryland 20742, USA*
[3] *BAE Systems, MER15-1813 P.O. Box 868, Nashua, New Hampshire 03061, USA*
[4] *Department of Engineering Physics, Air Force Institute of Technology, Wright-Patterson Air Force Base, Ohio 45433, USA*
[*]*alan.bristow@mail.wvu.edu*

**Abstract:** The angular dependence of terahertz (THz) emission from birefringent crystals can differ significantly from that of cubic crystals. Here we consider optical rectification in uniaxial birefringent materials, such as chalcopyrite crystals. The analysis is verified in (110)-cut $ZnGeP_2$ and compared to (zincblende) GaP. Although the crystals share the same nonzero second-order tensor elements, the birefringence in chalcopyrite crystals cause the pump pulse polarization to evolve as it propagates through the crystal, resulting in a drastically different angular dependence in chalcopyrite crystals. The analysis is extended to {012}- and {114}-cut chalcopyrite crystals and predicts more efficient conversion for the {114} crystal cut over the {012}- and {110}-cuts.

## 1. Introduction

Advances in broadband terahertz (THz) pulse generation by optical rectification have led to significant increases in available pulse energies, applicable for communications [1], imaging and spectroscopy [2], coherent control [3], and chemical recognition [4]. Efficient single-cycle pulses have been demonstrated by tilted-pulse-front pumping in LiNbO$_3$ [5], collinear phase matching in (110)-cut ZnGeP$_2$ (ZGP) [6,7], GaSe [8], and organic dimethyl amino 4-N-methylstilbazolium tosylate (DAST) [9] and 2-[3-(4-hydroxystyryl)-5,5-dimethylcyclohex-2-enylidene]malononitrile (OH1) [10] crystals, all of which are birefringent. In nonlinear optics in general, birefringence allows for flexibility of phase-matching conditions, but also adds potential complication due to the evolution of the polarization within the crystal. To make full use of various nonlinear optical crystals for broadband THz generation, it is useful to explore the effects of birefringence on optical rectification in uniaxial and biaxial materials.

An example of a positive-uniaxial birefringent THz source is ZGP [6,7], which is a chalcopyrite crystal and a ternary analog of the zincblende structure. Chalcopyrite crystals are of interest as THz sources due to their generally large nonlinearities and in some cases wide band gaps [11], the latter of which limits loss due to multiphoton absorption for near-infrared pumping. A previous comparative study [7] of the THz emission amplitude from ZGP, GaP, and GaAs showed ZGP to have suitable characteristics for excitation by near-infrared sources such as pulsed-fiber or chromium-doped forsterite lasers. However, it has a 2% lattice compression in the [001] direction, producing significant birefringence.

The work presented here explores the orientation dependence of uniaxial birefringent chalcopyrite crystals in order to optimize the phase-matching condition and build understanding of the optical rectification process. Chalcopyrite crystals provide an ideal choice for a study of the effects of birefringence due to their similarity to cubic zincblende structures. Because the second-order nonlinear tensors of these two structures have identical nonzero elements, differences in the respective angular dependence result either from differing values of the nonzero tensor elements or effects due to birefringence, and we argue here that the latter is dominant. First, analysis is presented comparing zincblende and chalcopyrite materials. Second, experimental verification of the analysis is presented. Finally, the analysis is extended to other crystal orientations to demonstrate optimization of the effective nonlinear coefficient.

## 2. Analysis

The second-order polarization due to optical rectification [12] is

$$P_i^{(2)} = \sum_{j,k} 2\varepsilon_o d_{ijk}(0,\omega,-\omega) E_j(\omega) E_k^*(\omega) \qquad (1)$$

where $\varepsilon_o$ is the permittivity of free space, $E_{j(k)}(\omega)$ are the pump electric fields and $d_{ijk}$ is the second-order nonlinear tensor. Zincblende crystals have $\bar{4}3m$ point group symmetry and three nonzero tensor elements, $d_{14}=d_{25}=d_{36}$. By comparison, chalcopyrite crystals have $\bar{4}2m$ symmetry and a similar tensor structure, with nonzero elements $d_{14}=d_{25}\neq d_{36}$ [13]. Typically, to obtain optimum optical rectification zincblende crystals are cut in the (110) plane, where the [001] direction is in the plane. The wavevector of the pump fields is normal to the crystal surface. The angle between the linear pump polarization and the [001] direction is defined as $\theta$. The emitted THz field can be determined as a function of $\theta$ for co-polarized ($E_{THz}^C$) and cross-polarized ($E_{THz}^X$) configurations, where in the former (latter) case the pump and THz polarizations are parallel (perpendicular).

Because ZGP is birefringent, the phase of the ordinary (o) wave, along $[\bar{1}10]$, evolves with respect to the phase of the extraordinary (e) wave, along [001]. Neglecting losses, the optical field at a depth $D$ within the crystal is

$$E(\theta,D,t) = E_o \sin\theta \exp[i(k_o D + \omega t)] + E_o \cos\theta \exp[i(k_e D + \omega t)], \quad (2)$$

where $E_o$ is the magnitude of the optical field, $\omega$ is the optical frequency, and $k_o$ and $k_e$ are the wavevectors of the ordinary and extraordinary components. For a pump field defined with arbitrary $\theta$ orientation Eq. (1) gives the nonlinear polarization in the crystal frame

$$P_X^{(2)}(D) = -2\sqrt{2}\varepsilon_o E_o^2 d_{14} \cos(k_d D)\cos\theta \sin\theta$$
$$P_Y^{(2)}(D) = 2\sqrt{2}\varepsilon_o E_o^2 d_{14} \cos(k_d D)\cos\theta \sin\theta \quad (3)$$
$$P_Z^{(2)}(D) = -2\varepsilon_o E_o^2 d_{36} \sin^2\theta$$

where the relative phase is defined as $k_d = k_e - k_o$. For $d_{14}=d_{36}$ and $k_d=0$, these expressions reduce to those of zincblende crystals [12,14]. For chalcopyrites however, the oscillating $\cos(k_d D)$ dependence of $P_X^{(2)}$ and $P_Y^{(2)}$ leads to generation of periodically interfering THz field components that become increasingly negligible for crystal lengths greater than $2\pi/k_d$. In contrast, the field components resulting from $P_Z^{(2)}$ increase linearly with the crystal length, assuming good velocity-matching of the optical and THz waves.

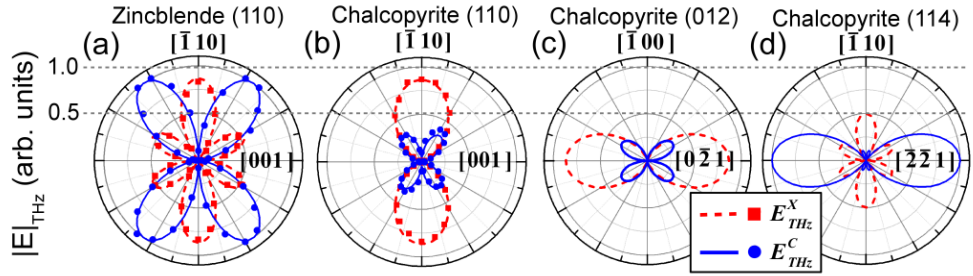

Fig. 1. Experimental and theoretical THz peak-to-peak field amplitude for a (a) (110)-cut GaP crystal with zincblende structure [14] and (b) (110)-cut ZnGeP$_2$ crystal with chalcopyrite structure and uniaxial birefringence. Theoretical prediction of the THz emission for uniaxial chalcopyrite structures cut in the (c) (012) plane and (d) (114) plane.

In (110)-cut ZGP, $2\pi/k_d \approx 30$ μm for 1300 nm excitation [15], which is a pump wavelength of interest. Typically, a real sample will be at least an order of magnitude longer than $2\pi/k_d$, and therefore, it is justified to neglect the $P_X^{(2)}$ and $P_Y^{(2)}$ completely, and the contribution from $d_{14}$ in this plane is effectively nonexistent. The co- and cross-polarized components of the THz emission are thus

$$E_{THz}^C(\theta) \propto P_C^{(2)} = -2\varepsilon_o E_o^2 d_{36} \cos\theta \sin^2\theta, \tag{4}$$

$$E_{THz}^X(\theta) \propto P_X^{(2)} = 2\varepsilon_o E_o^2 d_{36} \sin^3\theta. \tag{5}$$

Figure 1(a) and (b) show the prediction of $E_{THz}^C$ and $E_{THz}^X$ from (110)-cut zincblende and positive-uniaxial chalcopyrite structures as a function of the azimuthal angle. The two curves are on a relative scale, normalized to $(E_0)^2 d_{36}$, allowing for comparison of the THz field magnitudes. The maximum $E_{THz}^C$ component for chalcopyrite is reduced in comparison to the zincblende structure, but the maximum $E_{THz}^X$ component remains strong in both structures. This model is valid for crystal lengths shorter than the temporal walk-off length between o and e waves. The maximum effective $d$ coefficient is $d_{36}$ for chalcopyrite crystals, and $(2/3^{1/2})d_{36}$ for zincblende crystals.

## 3. Experimental Verification

The angular dependence of THz emission was measured in as-grown ZGP and undoped GaP samples with thicknesses of 0.33 mm and 0.1 mm respectively. As-grown ZGP single crystals were produced by the horizontal gradient freeze technique [16]. The ZGP sample was cut in the (110) plane and double-side polished for transmission measurements. The orientation of the crystals was verified by electron paramagnetic resonance, ensuring that the $\langle 001 \rangle$ direction is in the plane of the sample [17]. The transmission and birefringence homogeneity were verified with polarized transmission imaging. GaP is commercially available.

Measurements were performed at normal pump incidence with the crystal rotated about the axis of the pump wavevector. The generated THz field components were measured in directions parallel ($E_{THz}^C$) and perpendicular ($E_{THz}^X$) to the pump pulse polarization. Measurements employ ~100 fs pulses from a 1 kHz regenerative amplifier and optical parametric amplifier. For GaP and ZGP the pump pulses were tuned to center wavelengths of 800 nm and 1300 nm respectively. Electro-optic sampling, with linearly polarized 800 nm probe pulses and a (110)-cut ZnTe crystal, was used to detect the THz field. As expected, the measured spectra were broadband (0.5 THz to 3 THz), since the emission from GaP or ZGP is not limited by absorption due to IR-active phonons; see [7]. For ZGP, absorption in this frequency range is known to be smaller than that of ZnTe, GaP and and LiNbO$_3$ but similar to that of GaAs [18]. The electro-optic crystal detects one polarization component [19]. To acquire $E_{THz}^C$ and $E_{THz}^X$, the detection crystal and probe pulse polarization are rotated by 90°.

Field amplitudes are obtained by extracting the peak-to-peak voltage in individual THz transients as a function of $\theta$. The resulting angle-dependent field amplitudes for (110)-cut GaP and ZGP are shown in Fig. 1(a) and (b) respectively. The data exhibit excellent agreement with the model described above. ZGP is significantly longer than $2\pi/k_d$, hence the only surviving nonzero tensor element in the model is $d_{36}$. Consequently, discussion of the differences between the zincblende and chalcopyrite THz response resulting from the nonlinear tensor is moot, since the signal does not depend on $d_{14}$.

## 4. Predictions

The possibility of efficient generation from other crystal orientations of ZGP were explored using the above analytical approach which gives

$$E_{THz}^{C,(012)} \propto 2\varepsilon_0 E_0^2 d_{14} \cos^2\theta \sin\theta \tag{6}$$

$$E_{THz}^{X,(012)} \propto 2\varepsilon_0 E_0^2 d_{14} \cos^3\theta \tag{7}$$

$$E_{THz}^{C,(114)} \propto \left(2/6\sqrt{3}\right)\varepsilon_0 E_0^2 \cos\theta\left[8d_{14}\cos^2\theta + d_{36}(-1+5\cos 2\theta)\right] \tag{8}$$

$$E_{THz}^{X,(114)} \propto \left(2/6\sqrt{3}\right)\varepsilon_0 E_0^2 \sin\theta \left[8d_{14}\cos^2\theta + d_{36}(-1+5\cos 2\theta)\right] \quad (9)$$

for the C and X polarized field amplitudes in the (012) and (114) planes. Here, the azimuthal angle $\theta$ is generalized as the angle between the projection of [001] onto the plane of the crystal cut and the incident linear polarization direction.

Figure 1(c) and (d) show the predictions of the angle-dependent THz generation for uniaxial birefringent chalcopyrite crystals cut in the (012) and (114) planes respectively [Equivalent planes in zincblende crystals would be (011) and (112).] Each of these planes has a different near-infrared index of refraction for the e wave, but neither value varies as a function of $\theta$. Each crystal cut then exhibits unique phase matched wavelengths (not shown).

To understand the results it helps to inspect the crystal structures and orientations; see Fig. 2. For zincblende structures in the (110) plane the bonds are pointing $\pm 54.7^\circ$ away from the [001] projection; this is the angle at which $E_{THz}^C$ is maximized; see Fig. 1(a). For ZGP the angle is similar, if the 2% compression is ignored, but the birefringence strongly suppresses the THz generation for pump fields oriented at that angle. Reorienting the chalcopyrite to the (012) plane effectively rotates the bonds and the THz response by $90^\circ$ relative to the [001] projection, as seen by comparing Fig. 1(b) and (c). At the rotation angle for maximum efficiency in each of these planes the phase-matching is best achieved for cross-polarization, i.e. ooe for (110) cut and eeo for (012) cut crystals, where the first two waves correspond to the pump and the third corresponds to the emission. This can be considered as a relaxation of the singular phase matching condition observed in zincblende crystals.

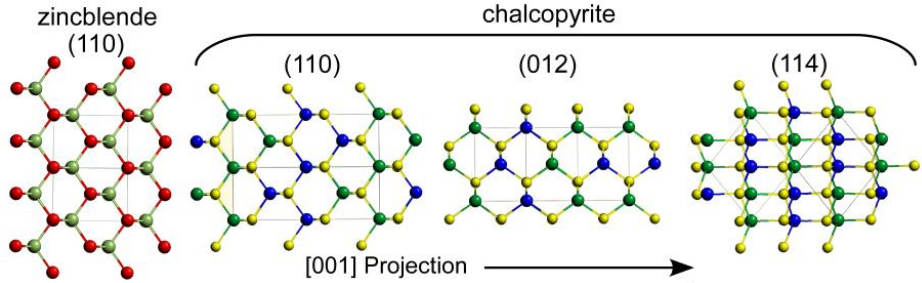

Fig. 2 Zincblende and chalcopyrite crystal structures for the planes considered. The arrow shows the direction of the [001] projected onto the plane of the page.

For chalcopyrite crystals, our analysis shows other efficient configurations. In fact, stronger emission should be observed for (114)-cut crystals. All of the angle dependences in Fig. 1 are plotted on the same relative scale, considering only the effects of birefringence with identical physical parameters; for a (114)-cut crystal, recovery of the maximum zincblende effective nonlinear coefficient $2/3^{1/2}d_{14}$ (assuming $d_{36}=d_{14}$) is predicted; see Fig. 1(d). For this orientation the crystal is rotated such that the bonds discussed above are now parallel to the [001] projection onto the crystal surface. The most efficient phase-matching condition is then eee, where all fields are collinearly polarized along this projection. For this orientation the crystal birefringence plays no role.

To extend the predictions to arbitrary crystal planes comprehensively, the analysis of Hargreaves *et al*. [14] for THz generation in zincblende crystals may be applied, taking into account the evolving pump polarization by numerically considering the crystal to be comprised of many thin slices. For a given $\theta$ the polarization state and orientation of the pump pulse are uniquely defined in the crystal frame by the relative phase between the o and e waves $\delta = k_d D$ [20], with $\beta$ defining the angle between the projection of the [001] direction and the major elliptical axis. Optical rectification is then modeled at each distance $D$ by considering field amplitudes along the major ($a$) and minor ($b$) axes of the elliptical polarization state also defined by $\delta$ at $\theta$.

This is illustrated for an arbitrary instantaneous elliptical polarization state in a (110)-cut crystal in Fig. 3(a). As *D* increases the polarization state evolves, oscillating between two linear polarization extremes about the optical axis of the crystal, as shown in Fig. 3(b) for various values of *δ*. At each position the major and minor components are used as individual instantaneous pump sources for THz generation within the analysis of [14]. Using the Hargreaves notation in this calculation, *β* is transformed to the lab frame angle Φ(*D*) with $\phi'=0$ and the incident pump polarization fixed along $\hat{z}''$. Then Eqn. (18) of [14] gives instantaneous polarizations resulting from the *a* or *b* field amplitudes, with $P_X(D) = P_{y''}^{(a)}(D) + P_{y''}^{(b)}(D)$ and $P_C(D) = P_{z''}^{(a)}(D) + P_{z''}^{(b)}(D)$. The $G_{ij}$ constants depend on the Miller indices of the cut plane, Φ (or *β*), $d_{36}$ and the incident field amplitude. A numerical sum of the field contributions at *D* for the full length of the crystal gives $E_{THz}^C$ and $E_{THz}^X$. The results of this model are in precise agreement with those shown in Fig. 1(a)-(d). This model gives a better physical picture of the fields and interactions inside the crystal.

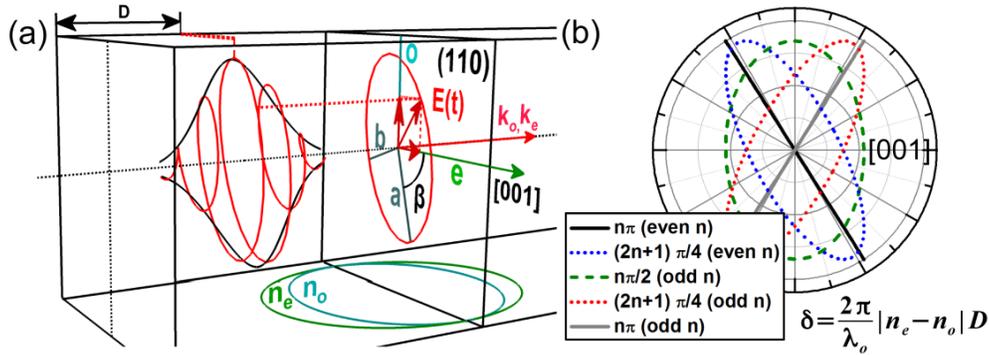

Fig. 3. (a) Arbitrary polarization state defined at a depth (*D*) in the crystal with uniaxial index ellipsoid. Major, minor axes and the rotation angle relative to [001] are *a*, *b* and *β* respectively. (b) Examples of the various pump polarization states at different relative phase delays in the crystal.

## 5. Conclusion

Previously, chalcopyrite crystals in the form of ZGP have been shown to be useful materials for THz generation by near infrared laser pulses [7]. The experimental and modeled results performed here show that birefringence modifies the angular dependence of the THz emission compared to the zincblende binary analogs. Birefringence is shown to reduce the maximum effective nonlinear coefficient for the commonly use (110) crystal orientation, which is often chosen based solely on the second-order tensor. In contrast it is predicted that (114) recovers the maximum effective nonlinear coefficient and the THz generation, because this orientation is not affected by birefringence.

This work highlights how birefringence plays a role in relaxing the phase-matching condition for THz generation in efficient emitters, without significant detrition in the efficiency. The model can easily be extended to other birefringent materials, given the appropriate second-order nonlinear optical susceptibility tensor.

**Acknowledgements**

The authors wish to thank Larry Halliburton and John Sipe for useful discussions. JDR wishes to thank the WVNano Initiative for support. The views expressed in this article are those of the authors and do not necessarily reflect the official policy or position of the Air Force, the Department of Defense, or the United States Government.